\documentclass[preprint2]{aastex}
\usepackage{graphicx}
\usepackage{amssymb}
\usepackage{epstopdf}
\usepackage{IEEEtrantools}
\usepackage{natbib}

\newcommand{\beq}{\begin{equation}}
\newcommand{\eeq}{\end{equation}}
\newcommand{\beqa}{\begin{eqnarray}}
\newcommand{\eeqa}{\end{eqnarray}}
\newcommand{\pp}{^{\prime\prime}}
\newcommand{\kms}{\ {\rm km\,s^{-1}}}
\newcommand{\msun}{M_{\odot}}
\newcommand{\etal}{{\it et al.\ }}

\shorttitle{The mass of the MW and M31}
\shortauthors{Phelps et al.}

\begin{document}
\title{The mass of the Milky Way and M31 using the method of least action}

\author{Steven Phelps}
\affil{Physics Department, Technion, Haifa 32000, Israel}
\email{steven@alumni.princeton.edu}

\author{Adi Nusser}
\affil{Physics Department and the Asher Space Science Institute-Technion, Haifa 32000, Israel}
\email{adi@physics.technion.ac.il}

\and

\author{Vincent Desjacques}
\affil{Departement de Physique Theorique and Center for Astroparticle Physics, Universite de
Geneva, CH-1211 Geneva, Switzerland}
\email{Vincent.Desjacques@unige.ch}

\begin{abstract}

We constrain the most likely range of masses for the Milky Way (MW) and M31 using an application of the numerical action method (NAM) that optimizes the fit to observed parameters over a large ensemble of NAM-generated solutions.  Our $95\%$ confidence level mass ranges, $1.5 - 4.5 \times 10^{12} \msun$ for MW and $1.5 - 5.5 \times 10^{12} \msun$ for M31, are consistent with the upper range of estimates from other methods and suggests that a larger proportion of the total mass becomes detectable when the peculiar motions of many nearby satellites are taken into account in the dynamical analysis.  We test the method against simulated Local Group catalogs extracted from the Millennium Run to confirm that mass predictions are consistent with actual galaxy halo masses.

\end{abstract}

\keywords{cosmology: theory -- galaxies : kinematics and dynamics -- galaxies:  Local Group}

\section{Introduction}

Estimating the total masses of galaxies, our own in particular, is a continuing challenge of precision cosmology.  Part of the challenge lies in the unknown extent of the dark matter halos within which they are presumably embedded:  while the measurement of galaxy rotation curves from coherent stellar motions allows the mass within the visible radius to be inferred, the total mass of the associated dark matter halos predicted in the standard model of cosmology, whose physical extent is not known, is more difficult to estimate.  To probe the total effective gravitational mass the analysis must include the effect on the peculiar motions of nearby galaxies. 

The measurement of total galaxy masses from their relative motions was pioneered by Kahn \& Woltjer (1959).  Their ``timing argument" (TA) method, which assumes purely radial infall, indicated a total mass for the MW+M31 system of about $3 \times 10^{12} \msun$---a lower bound, since the possibility of transverse motions is excluded.  The total mass of the Local Group (LG) can also be computed from the velocity dispersion of its various member galaxies, assuming that it is in virial equilibrium and that its velocity ellipsoid is isotropic:  Courteau \& van den Bergh (1999), using this method, found a LG mass of $(2.3 \pm 0.6) \times 10^{12} \msun$.  More recent applications of the TA tend to suggest higher masses.  Li \& White (2008) confirmed that the TA method used on mock galaxies drawn from the Millennium Run (Springel \etal 2005) systematically underestimates the true mass, and revised the TA method to predict a LG mass of $(5.27 \pm 0.5) \times 10^{12} \msun$.  Van der Marel \etal (2012) used the full proper motion of M31 to improve the TA method, estimating a LG mass of about $5 \times 10^{12} \msun$, somewhat higher than the combined prediction of $(3.17 \pm 0.57) \times 10^{12} \msun$ from a Bayesian combination of estimates from different methods.  The TA can also be used with the proper motion of Leo I (Sohn \etal 2013), assuming it is gravitationally bound to the Milky Way (MW), to estimate the mass of the MW alone.  Boylan-Kolchin \etal (2013) combine the TA and other methods in estimating a virial mass for the MW of $1.6 \times 10^{12} \msun$ with a 90\% confidence interval of $[1.0 - 2.4] \times 10^{12} \msun$.  

Mass estimation methods using the TA are based on the analysis of single-galaxy interactions with the Milky Way.  We show in this paper that the numerical action method (NAM), by taking into account the peculiar motions of a large subset of Local Group satellites, effectively breaks the mass degeneracy in the two-body TA and identifies separate ranges of likely masses for the two principal actors in the LG.  The method avoids the TA assumption that galaxies are gravitationally bound, makes no assumptions about virialization of the LG, and is sensitive to more widely diffused concentrations of dark matter that could remain undetected using other methods.  NAM takes as input the cosmological parameters $H_0$ and $\Omega_0$, and assumes that linear theory correctly describes velocities at early times and that galaxies and their progenitors back in time can be approximated as simple paths representing the center-of-mass motions of their associated dark matter halos.

Earlier papers on NAM (including Peebles 1989 introducing the method; Peebles 1995; Peebles, \etal 2001; Phelps 2002; Peebles, \etal 2011; Peebles \& Tully 2013) developed the method in the context of the Local Group, focusing on the analysis of individual solutions or small ensembles.  In what follows we explore the behavior of several thousand independent solutions in order to identify the combinations of MW and M31 masses that yield the best fit to the observational constraints.  Section 1 introduces our implementation of NAM.  In Section 2 we apply it to the Local Group, and in Section 3 we test NAM predictions in the LG with mock catalogs drawn from the Millennium Run.  A brief discussion including an assessment of future prospects is found in Section 4.

\section{The numerical action method}

\subsection{Method and Approximations}

Our version of NAM is based on the improved algorithm described in Peebles \etal (2011).  We have extended it to include an optional partial canonical transformation of coordinates between the radial distance and the radial velocity (first used in Peebles \etal 2001 and more fully described in Phelps 2002), allowing either the galaxy distance or redshift to be chosen as the boundary condition in the action integration at $z=0$, in addition to the two components of the observed galaxy position in the plane of the sky.  This effectively doubles the solution space that can be explored.  The Appendix gives details of the coordinate transformation.

We model galaxies as constant-density spheres with MW radius of 100 kpc and all other galaxy radii scaling proportionately with the cube root of their masses.  From the start of the computation time at $a=0.1$ until $a = 1$, galaxy paths are interpreted as the mean motions of the coalescing systems of baryonic matter and their associated dark matter halos.  As shown in Peebles \etal (2011), the initial velocities are by construction consistent with linear perturbation theory.  This keeps the galaxies well separated from each other at early times, and so passages of galaxies through each other's (nonphysical) cutoff radii are rare.

Galaxy flight paths are reconstructed from initial randomized straight line trial orbits by successive iterations in the direction of the first and second derivatives of the action until a stationary point is reached.  Solutions are verified by comparing them to the solutions to the equations of motion from the same initial timestep in a leapfrog approximation.  With a target of $10^{-11}$ in the sum of squares of the gradient of the action, deviations between NAM-generated flight paths and the leapfrog approximation are at most a few kpc at the final timestep.  In the present work we follow Peebles \& Tully (2013) in excluding several close satellites to the MW, but we use a smaller number of time steps (30, versus 500 in Peebles \etal 2011 and Peebles \& Tully 2013), which is still sufficient to produce tightly bending half-orbits.  This significantly reduces computation time and maximizes the number of solutions generated.  

NAM solutions are non-unique owing to the mixed boundary conditions:  velocities are constrained at the initial time (see discussion in the Appendix, A.1.4), while some combination of distances and velocities are fixed at the final time.  As a consequence, different choices of initial trial orbits with the same observational constraints will in general yield a different set of galaxy paths, each of which are solutions to the equations of motion.  However, as there are more constraints than those required for a solution (it is sufficient to specify either distances or redshifts), a $\chi^2$ measure of fit is defined for each observable,
\beq
\chi^2 = \sum_{i} \left({model_i - cat_i \over \sigma_i}\right)^2 
\eeq
from which a best-fit solution can be selected from an ensemble.

Apart from the initial trial orbits, solutions may also be sensitive functions of the input parameters, particularly for galaxies in close proximity to each other.  These parameters are fixed in the formal NAM computation even though there is an observational uncertainty associated with each.  Since we wish to explore the widest possible range of reasonable physical configurations, we adopt the method, similar to Peebles \etal (2011) and Peebles \& Tully (2013), of defining a $\chi^2$ measure as a sum over all constraints (distance, redshift, angular position, mass, and, where available, velocity transverse to the line of sight), fixing the observational constraints as their given values plus a random error within the observational uncertainty that differs for each trial solution. Our per-particle $\chi^2$ is defined as follows:
\beqa
\chi^2_{tot} & = & \sum_{i} ( \chi_{i,d}^2 +\chi_{i,cz}^2 + \chi_{i,\theta}^2 + \chi_{i,\phi}^2 \nonumber \\
 & & + \chi_{i,mass}^2 + \chi_{i,vtrv}^2 + \chi_{i,v0}^2 )
\eeqa
We relax the initial trial orbits to a solution to the equations of motion using NAM, and then relax each NAM solution to a minimum in $\chi^2$.  By holding some of the observational constraints relatively fixed, we can repeat this procedure for a large number of NAM solutions to explore an $n$-dimensional space of solutions whose minimum in $\chi^2$ gives the most likely values of the desired $n$ constraints.  In our case we are interested in the masses of the principal actors in the Local Group and so we are looking for a minimum within the two-dimensional space defined by the masses of MW and M31.  Further details of our minimization approach, which is fully general and can be applied to any desired combination of observational parameters, are as follows.

For each solution we assign masses to MW and M31 within a range of $0.5$ and $6 \times 10^{12}\msun$ in a Gaussian random distribution centered around the nominal masses given in the catalog.  This produces higher quality solutions by focusing less attention on combinations of masses which experience has shown are less likely to yield good fits to the constraints.  Random errors are added to all other observables (distances, redshifts, angular positions, and masses), with standard deviations taken generically to be 10\% of the distance for distances, $5 \kms$ for redshifts, 60\% of the nominal catalog masses (giving these quantities the widest possible latitude), half a degree for angular positions, and the published uncertainties for proper motions (as summarized in Peebles and Tully 2013; these are available for M31, LMC, M33, IC10, and LeoI).  Our relative leniency in angular positions, which we allow owing to the approximate nature of the model, also accounts for the physical possibility that galaxies are offset from the center of mass of their respective dark matter halos.  Since the main objective of this study is investigating the masses of MW and M31, and good agreement with M31 constraints is therefore a priority, standard deviations in distances, redshifts, and positions for galaxies beyond 1.5 Mpc are doubled and those for M31 are halved.  Standard deviations in MW and M31 masses are reduced by a factor of 20 relative to their initial randomized guesses, essentially fixing them in the $\chi^2$ relaxation step but allowing them to shift by a small amount if a significantly better solution is found at a slightly different mass.  As large initial velocities are permitted in the solutions (see Appendix A.1.4), a contribution to $\chi^2$ from the magnitude of the initial velocity (the final term in Equation (2) above) is additionally assigned with a standard deviation of $40 \kms$.  This effectively suppresses implausibly large initial velocities in the search for the best solutions.

We find that NAM solutions are efficiently generated by generating a solution for the two dominant galaxies first and then adding the other galaxies, one at a time, in descending order of mass.  With each NAM solution we then take each galaxy in turn in the same order, using Powell's method to relax the individual galaxy distance, redshift and angular position to a minimum in its $\chi^2$ since derivatives in $\chi^2$ with respect to these quantities cannot be reliably computed.  Discontinuities in $\chi^2$ may be encountered where a galaxy jumps to a qualitatively different orbit in the descent to the target in the action; these are allowed so long as it leads to an improvement in $\chi^2$.  If after relaxing with Powell's method $\chi^2$ remains above a given threshold (100 yielded a good balance between time to solution and quality of solution), we recast the orbit for this galaxy and find another solution.  If, after 50 attempts, $\chi^2$ is still above the threshold, we switch from distance to redshift boundary conditions for this galaxy and allow another 25 attempts.  If $\chi^2$ still remains above the threshold, we take the solution corresponding to the lowest $\chi^2$ found thus far and move to the next galaxy.  In practice between 25\% and 50\% of the galaxies in any given NAM solution will have been fitted to the velocity boundary condition and the rest to the distance boundary condition.  Once $\chi^2$ has been minimized in this way, we then jointly relax the galaxy masses, also using Powell's method, to minimize $\chi^2$ still further (we found that relaxing the masses separately from the other quantities gave more rapid convergence to a minimum).  This method of relaxation between the various observational quantities typically reduces $\chi^2$ by an order of magnitude from its initial value; further improvements to this ad hoc procedure are no doubt possible.

We generate four thousand independent solutions using this method, varying the mass of MW and M31 in each solution as described above.  Parallelizing the code to run on 12 supercomputing nodes using 2.40 GHz six-core Xeon processors, each solution typically takes 10 or 20 seconds for catalogs up to a dozen or two particles.  From an ensemble of solutions we then plot a Gaussian-smoothed $\chi^2$ map against the MW and M31 masses, dividing the map into $24 \times 24$ equal bins and keeping the best $\chi^2$ in each bin.  The smoothing is desirable because the solution space can feature many local minima, each corresponding to a qualitatively different configurations of orbits. 

\section{Results for the Local Group}

Our Local Group catalog, based on Peebles \& Tully (2013), is listed in Table 1.  The omission of the smallest, tightly bound satellites of MW and M31 facilitates direct comparison to the simulations, which lack comparable range in mass, and maximizes the number of solutions which can explored.  The four actors listed immediately after MW and M31 are intended to approximately model the influence of the significant mass concentrations residing immediately beyond the Local Group.  While inclusion of all Local Group actors may hold the promise of producing even better dynamical constraints on the mass of MW and M31, it is offset by the greatly increased effort required to produce acceptable solutions.

\begin{table}

\caption{The Local Group Catalog: Observational Constraints}
\smallskip
\smallskip 
\begin{tabular}{lllllll}

\hline\hline\noalign{\smallskip}
Name  &   d  &  SGL &     SGB    &   cz   &  $m_{cat}$  \\ \hline\noalign{\smallskip}
MW  &  0.00 &   0.00 &   0.00 &     0 &   22.5 \\
M31  &  0.79 & 336.19 &  12.55 &  -119 &   25.1 \\ 
Cen+  &  3.57 & 159.75 &  -5.25 &   386 &  119.0 \\
M81+  &  3.66 &  41.12 &   0.59 &    72 &   70.6 \\
Maff+  &  3.61 & 359.29 &   1.44 &   168 &   70.2 \\
Scp+  &  3.58 & 271.56 &  -5.01 &   251 &   50.6 \\
M33  &  0.92 & 328.47 &  -0.09 &   -45 &    1.97 \\
LMC  &  0.05 & 215.80 & -34.12 &    66 &    1.24 \\
IC10  &  0.79 & 354.42 &  17.87 &  -150 &    0.434 \\
NGC185  &  0.64 & 343.27 &  14.30 &   -40 &    0.106 \\
NGC147  &  0.73 & 343.32 &  15.27 &    -4 &    0.077 \\
NGC6822  &  0.51 & 229.08 &  57.09 &    43 &    0.06 \\
LeoI  &  0.26 &  88.90 & -34.56 &   121 &    0.001 \\
LeoT  &  0.16 &  78.40 & -38.75 &   -61 &    0.001 \\
Phx  &  0.43 & 254.29 & -20.86 &   -34 &    0.001 \\
LGS3  &  0.65 & 318.13 &   3.81 &  -149 &    0.001 \\
CetdSph  &  0.73 & 283.84 &   3.82 &   -27 &    0.001 \\
LeoA  &  0.74 &  69.91 & -25.80 &   -13 &    0.001 \\
IC1613  &  0.75 & 299.17 &  -1.80 &  -159 &    0.001  \\ \hline\noalign{\smallskip}     
\end{tabular}

Units:  Mpc, deg, $\kms$, $10^{11}\msun$.

\end{table}

$\chi^2$ maps for 4000 solutions in four different scenarios, all with $H_0 = 67$ and $\Omega_0 = 0.27$, are shown in Figure 1.   At upper left are the contours in $\chi^2$ generated from a simplified catalog consisting of only MW and M31, to check the consistency of our implementation of NAM with the Timing Argument.  As expected from the classic TA, we find a well-defined constraint on the sum $m_{MW} + m_{M31}$ but are unable to resolve the individual masses.  The bare TA sum indicated by NAM, ($m_{LG} = 6 \pm 1) \times 10^{12}\msun$ at $95\%$ confidence, is consistent with Li and White (2008), who calibrated the TA against galaxy pairs drawn from the Millennium simulation to find, at $95\%$ confidence, $1.9 \times 10^{12} \msun < M_{LG} < 1.0 \times 10^{13} \msun$, with a median likelihood estimate of $5.7 \times 10^{12} \msun$ (the confidence interval in the latter is larger since it takes cosmic variance into account).

In Figure 1, upper right, we show the results from a reduced version of our catalog which includes the LG actors but excludes the four external groups.  The additional of the additional dynamical actors has broken the degeneracy in the TA, giving independent masses of $2.5 \pm 1.5 \times 10^{12} \msun$ for the MW and $3.5 \pm 1.0 \times 10^{12} \msun$ for M31.  With the addition of the four external groups (Figure 1, lower left), the best mass for the MW increases to $3.5 \pm 1.0 \times 10^{12} \msun$.  This is consistent at the lower end with previous TA measurements of the total LG mass and the individual MW mass.  When the transverse velocity constraints on M31, LMC, M33, IC10, and LeoI are added (Figure 1, lower right), the confidence intervals are broadened and the best-fit mass for MW decreases slightly, to $3.0 \pm 1.5 \times 10^{12} \msun$, reflecting the fact that lower masses for MW are correlated to lower transverse velocities for M31 and other nearby galaxies. 

\begin{figure}
\includegraphics[width=3.5 in.]{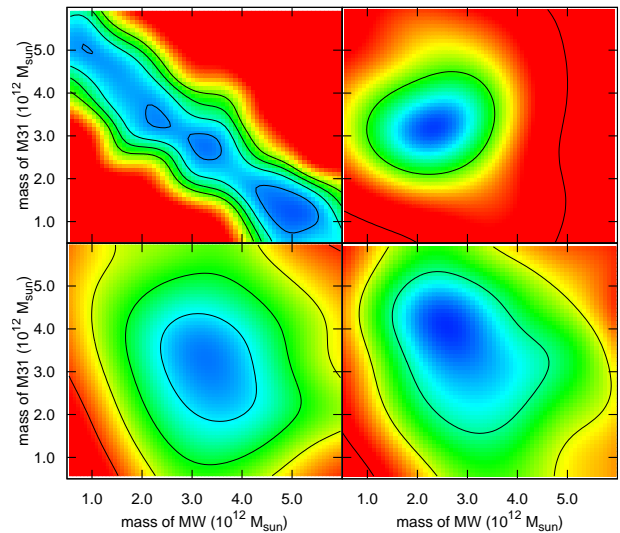}
\caption{Contours in $\chi^2$ for different values of MW (x-axis) and M31 (y-axis) masses, for 4000 NAM solutions in four different scenarios.  Upper left: results from the two-body problem of MW + M31.  Upper right: LG actors only.  Lower left: LG actors + four external groups.  Lower right: same as lower left, with transverse velocity constraints included for five nearby galaxies.  The first contour level (solid black line) marks the region of 95\% confidence (minimum $\chi^2$ + 6 for two degrees of freedom).}
\label{fig:figure1}
\end{figure}

\section{A test of NAM mass predictions in simulations}

As a check on the result for the Local Group, we use publicly available data from the Millennium Run (Springel \etal 2005) and follow Li \& White (2008) in generating mock Local Group catalogs satisfying the following conditions:  We select type 0 or 1 galaxies with rotation velocities in the range $200 < V_{max} < 250 \kms$ and bulge-to-total luminosity ratio in the range $1.2-2.5$.  We then sub-select close pairs with comoving separation between 0.5 and 1 Mpc, negative relative peculiar velocity and no massive companion with $V_{max} > 150 \kms$ within 2.5 Mpc from the center-of-mass. This led to the initial identification of about 100 Local Group candidates.  Since the simulation overproduces satellite galaxies relative to observations (the ``missing satellite problem"; see, e.g., Bullock 2012), and since the resulting dynamical complexity poses a special difficulty for NAM reconstructions, we additionally exclude catalogs where the gravitational acceleration on mock-MW due to satellites is greater by more than a factor of six than what we expect from the observed distribution (assuming the distance and nominal masses listed in our catalog).  This limited our set of mock catalogs to 32 with reasonably similar dynamics to what expect to be the case with the LG, although in every case but two the dynamical complexity as defined above is greater than in the LG.  To speed up computation time, within each of the 32 selected mock catalogs we include in the NAM computation only those galaxies and satellites within the distance to mock-M31, and all other galaxies out to 7 Mpc which produce an acceleration of at least 5\% that of M31.  This reduced the number of particles to those that are most relevant to the dynamics of the principal actors -- from as few as 9 to as many as 34, with an average across the catalogs of 19.

Contours in $\chi^2$ for different halo masses of mock-MW and mock-M31 are shown for the four dynamically simplest catalogs in Figure 2.  The predicted values for their masses are consistent with the true galaxy + halo masses from the simulation, within the $95\%$ confidence range.  We may expect the NAM prediction to be somewhat higher if it is sensitive to dark matter associated with the two principal actors but beyond their respective cutoff radii (at an overdensity of 200) that define their masses.  A detailed comparison of the underlying dark matter distribution with the halo mass predictions may indicate whether this sensitivity is present.  For the other 28 mock catalogs the mass predictions for both actors were likewise consistent with the $95\%$ confidence intervals in all but three cases. The NAM-based mass predictions from the mock catalogs are thus accurate within statistical expectations.

\begin{figure}
\includegraphics[width=3.5 in.]{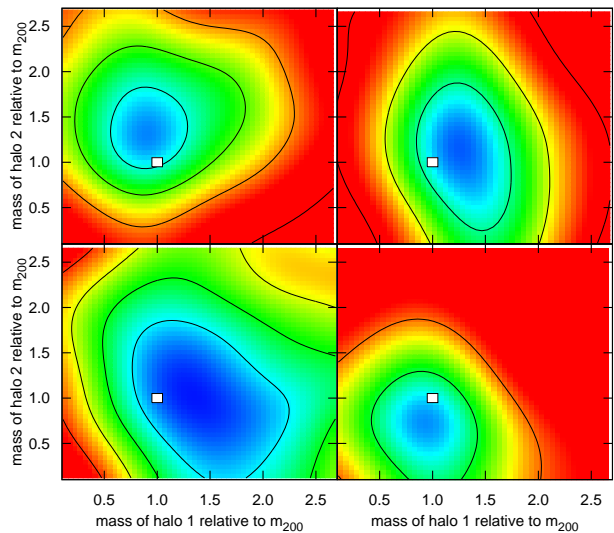}
\caption{Contours in $\chi^2$ for four different simulated Local Group catalogs. Each plot is generated from 4000 independent NAM solutions.  Axes are unitless and represent the ratio of the NAM mass to the actual halo mass in the simulation; thus the point at (1.0, 1.0), marked with small squares, corresponds to a NAM solution using the actual halo masses.  The first contour level (solid black line) marks the region of 95\% confidence (minimum $\chi^2$ + 6).}
\label{fig:figure2}
\end{figure}

\section{Discussion}

We have shown that the Numerical Action Method offers a promising method of constraining individual masses of the principal actors in the Local Group using an approach that investigates the behavior of $\chi^2$ across a large ensemble of solutions.  It is complementary to earlier work focusing on the detailed dynamical analysis of individual solutions.  

The relatively large masses of MW and M31 suggested in this first-of-its-kind implementation of NAM are a point of interest.  They may be due to the method's sensitivity to both the extended dark matter halo as well as purely dark concentrations between galaxies, since these otherwise invisible concentrations should leave a signature in galaxy proper motions.  In this case, our result would suggest the presence of mass concentrations larger than previously suspected.  Another possibility is that the result is biased by poor reconstruction of satellite paths, since solutions placing M31 at exactly the observed distance and redshift also tend to place the dwarf galaxies in the near vicinity of the MW at distances greater than the catalog distance.  Related to the above point, our observational constraints assume a value of 220 km/s for the circular velocity of the sun around the MW center; a higher value would reduce the approach velocity of M31, reducing the predicted mass.

Further improvements to this method are suggested by the above observations, and can be expected from a number of directions.  On the computational side, increases in the efficiency of the solution finding algorithm will permit a larger number of time steps to be used and potentially allow for more complex orbits, improving in particular the reconstruction of nearby satellites.  It will also permit a larger complement of dwarf and satellite galaxies to be included.  Adding full three-dimensional proper motions of a larger number of nearby galaxies will certainly further constrain the likely masses.  A comparison of the underlying dark matter distribution with the galaxy halos in the simulations will confirm whether NAM is potentially sensitive to extended distributions of dark matter beyond the nominal halo radii.  

We thank Brent Tully for helpful suggestions that improved the presentation of the paper.  This research project was supported at the Technion by the I-CORE Program of the Planning and Budgeting Committee and the ISF.  V. Desjacques acknowledges support by the Swiss National Science Foundation.


\appendix
\section{Appendix:  A faster NAM in redshift space}

Sections A.1 and A.2 are from Peebles \etal (2011).  Section A.3 is new to this study and shows how the action can be modified to accommodate redshift boundary conditions.

A.1. {\bf Review of the theory}
 
 A.1.1. {\it Equations of motion}

\smallskip

In a cosmologically flat universe the expansion parameter satisfies 
\beq
{\dot a^2\over a^2} = {H_o^2\Omega\over a^3} + (1 - \Omega)H_o^2,\qquad
{\ddot a\over a} = -{H_o^2\Omega\over 2a^3} + (1 - \Omega)H_o^2,
\eeq
with present value $a_o=1$. The equations of motion in physical length units are
\beq
{d^2r_{i,k}\over dt^2} = \sum_{j\not= i} {Gm_j(r_{j,k}-r_{i,k})\over |r_i-r_j|^3} +(1-\Omega)H_o^2r_{i,k}.\label{eq:2}
\eeq
Changing variables to the comoving coordinates $x_{i,k}=r_{i,k}/a(t)$  used here brings eq. (\ref{eq:2}) to 
\beq
{d\over dt}a^2{d\,x_{i,k}\over dt}= {1\over a}\left[ \sum_{j} Gm_j{(x_{j,k}-x_{i,k})\over |x_i-x_j|^3} +{1\over 2}\Omega H_o^2x_{i,k}\right].
\label{eq:3}
\eeq
This is derived from the action 
\begin{IEEEeqnarray}{rCl}
S & = & \int_0^{t_{o}}dt\left[\sum_i {m_i a^2 \dot x_i^2\over 2} +
{1\over a}\left(\sum_{j\not= i}{Gm_im_j\over |x_i-x_j|} 
+ {1\over 4}\sum_i m_i\Omega H_o^2 x_i^2\right)\right] \label{eq:99}
\end{IEEEeqnarray}
when the present positions are fixed, $\delta x_i(t_o)=0$, and initial conditions satisfy
\beq
a^2\dot x_i\rightarrow 0 \hbox{ at } a(t)\rightarrow 0.
\label{eq:5}
\eeq

 \medskip
A.1.2. {\it Discrete representation}

In a discrete representation the coordinates are  $x_{i,k,n}$, where $i$ labels the particles, $k=1,2,3$ the Cartesian coordinates, and $1\leq n\leq n_x+1$ the time steps. The present positions $x_{i,k,n_x+1}$ are fixed and given.  The relevant derivatives of the action are 
\beq
S_{i,k,n}={\partial S\over\partial x_{i,k,n}}, \quad
S_{i,k,n;j,k',n'}={\partial^2 S\over\partial x_{i,k,n}\partial x_{j,k',n'}},
\quad 1\leq n,n'\leq n_x. \label{eq:8}
\eeq
If $S$ is close to quadratic in the $x_{i,k,n}$ then position shifts $\delta x_{i,k,n}$ to a solution at an extremum of $S$ satisfy
\beq
S_{i,k,n} + \sum_{j,k',n'}S_{i,k,n;j,k',n'}\delta x_{j,k',n'} =0. \label{eq:shifts}
\eeq
If the $x_{i,k,n}$ are not close to a solution $S$ is not close to quadratic in the $x_{i,k,n}$, but experience shows that coordinate shifts in the direction of  $\delta x_{i,k,n}$ walk toward a solution. 

Approximate the action (\ref{eq:99}) as
\beqa
S &=& \sum_{i,k,n=1,n_x}{m_i\over 2}{(x_{i,k,n+1}- x_{i,k,n})^2\over (a_{n+1}-a_n)}
\dot a_{n+1/2}a_{n+1/2}^2 \nonumber\\
&+&\sum_{i,j,n=1,n_x}{t_{n+1/2}-t_{n-1/2}\over a_n}
\left[\sum_{j<i}{Gm_im_j\over  |x_{i,n}-x_{j,n}|} + {1\over 4}
\sum_i m_i\Omega H_o^2 x_{i,n}^2\right]. \label{eq:9}
\eeqa
The times $t_{n\pm 1/2}$ interpolate between the time steps at $n$ and $n\pm 1$ in leapfrog fashion.  The approximation to the kinetic energy in eq.~(\ref{eq:9}) is motivated by linear perturbation theory, where $dx/da$ is nearly independent of time, so $(x_{n+1}- x_{n})/(a_{n+1}-a_n)$ is a good approximation to $dx/da$ at $a_{n+1/2}$. The earliest time at which positions are computed is at $a_1>0$. The leapfrog back in time from $a_1$ is to $a_{1/2}=0=t_{1/2}$. Recall that present positions at $a_{n_x+1}=1$ are given at $x_{i,k,n_x+1}$. 

The derivative of the action with respect to the coordinates $x_{i,k,n}$ for $1\leq n\leq n_x$, gives 
\beqa
S_{i,k,n} &=& -{a_{n+1/2}^2\dot a_{n+1/2}\over a_{n+1} - a_n}(x_{i,k,n+1}-x_{i,k,n}) \nonumber\\
&+& {t_{n+1/2}-t_{n-1/2}\over a_n}
\left[\sum_{j\not=i}Gm_j{x_{j,k,n}-x_{i,k,n}\over |x_{i,n}-x_{j,n}|^3} +
{1\over 2}\Omega H_o^2 x_{i,k,n}\right]\label{eq:10}\\
&+& {a_{n-1/2}^2\dot a_{n-1/2}\over a_{n} - a_{n-1}}(x_{i,k,n}-x_{i,k,n-1}). \nonumber
\eeqa
When $S_{i,k,n}=0$ this is a discrete approximation to the equation of motion~(\ref{eq:3}). The common factor $m_i$ has been dropped to reduce clutter, which means $S_{i,k,n;j,k',n'}\not= S_{j,k',n';i,k,n}$. (The asymmetry is in the gravity term. There still is the symmetry $S_{i,k,n;i,k',n'} = S_{i,k',n';i,k,n}$.) 

To simplify eq. (\ref{eq:10}) and its derivatives wrt $x_{i,k,n}$ let
\beq
F^+_n=  
{a_{n+1/2}^2\dot a_{n+1/2}\over a_{n+1} - a_n}, 
\quad
 F^-_n=  
{a_{n-1/2}^2\dot a_{n-1/2}\over a_{n} - a_{n-1}}
= F^+_{n-1}, \quad {dt_n\over a_n} = {t_{n+1/2}-t_{n-1/2}\over a_n}.
\label{eq:F}
\eeq
Note that 
\beq
F^-_1 = 0=F^+_0  \label{eq:bdyconds}
\eeq
follows from $a^2\dot a\rightarrow 0$ at $a\rightarrow 0$. Also, write the acceleration (apart from the factor $a^2$) of particle $i$ due to the other particles $j\not= i$ as 
\beq
g_{i,k,n} = \sum_{j\not= i} Gm_j{x_{j,k,n}-x_{i,k,n}\over |x_{i,n}-x_{j,n}|^3}. 
\eeq
All this notation brings eq. (\ref{eq:10}) to
\beq
S_{i,k,n} = -F^+_n(x_{i,k,n+1}-x_{i,k,n}) +  F^-_n(x_{i,k,n}-x_{i,k,n-1})
+ {dt_n\over a_n}\left[g_{i,k,n}  + 
{1\over 2}\Omega H_o^2 x_{i,k,n}\right].\label{eq:16}
\eeq

We need the derivatives of the action with respect to the positions of the particles. Let the derivative of the acceleration of particle $i$ wrt the position of particle $j\not= i$ be
\beqa
{\cal G}_{i,k,n;j,k'} &=&  {\partial g_{i,k,n}\over\partial x_{j,k',n}} \nonumber\\
&=& Gm_j\left(  {\delta_{k,k'}\over |x_{i,n} - x_{j,n}|^3} -
3{(x_{j,k,n}-x_{i,k,n})(x_{j,k',n}-x_{i,k',n}) \over 
|x_{i,n} - x_{j,n}|^5 }\right). 
\eeqa
The derivative of the acceleration of particle $i$ with respect to its own position is 
\beq
{\cal G}_{i,k,n;i,k'} =  {\partial g_{i,k,n}\over\partial x_{i,k',n}} =
 -\sum_{j\not= i} {\cal G}_{i,k,n;j,k'}
\eeq
So the nonzero derivatives of eq. (\ref{eq:16}) with respect to the coordinates are
\beqa
S_{i,k,n;i,k,n+1} &=& - F^+_n, \qquad S_{i,k,n;i,k,n-1} = - F^-_n, \nonumber\\
 S_{i,k,n;j,k',n} &=& {dt_n\over a_n} {\cal G}_{i,k,n;j,k'}, \ j\not= i, 
 \label{secondderivatives} \\
 S_{i,k,n;i,k',n}&=& (F^+_n + F^-_n)\delta_{k,k'} +  
{dt_n\over a_n}\left[ {\cal G}_{i,k,n;i,k'}
+{1\over 2}\Omega H_o^2\delta_{k,k'}
 \right].\nonumber
\eeqa
The goal is to use these second derivatives of the action to drive the first derivatives to zero at a stationary point, $S_{i,k,n}=0$. 

\bigskip

A.1.3. {\it Leapfrog integration forward in time}

It is worth recording that when the action is at a stationary point, $S_{i,k,n}=0$, a solution of equation~(\ref{eq:16}) is equivalent to a standard leapfrog numerical integration of the equation of motion (\ref{eq:3}) forward in time. In this leapfrog, the positions and velocities are computed at interleaved time steps as (in the notation in eq [\ref{eq:F}]),
\beq
a_{n+1/2}^2\dot x_{i,k,n+1/2} = a_{n-1/2}^2\dot x_{i,k,n-1/2} + {dt_n\over a_n}\left[\sum_{j\not= i}g_{i,k,n;j}  + 
{1\over 2}\Omega H_o^2 x_{i,k,n}\right],\label{eq:fwd1}
\eeq
and\beq
 x_{i,k,n+1} =  x_{i,k,n} + \dot x_{i,k,n+1/2}(t_{n+1}-t_n)=
x_{i,k,n} + \dot x_{i,k,n+1/2}(a_{n+1}-a_n)/\dot a_{n+1/2},
\eeq
which in the notation in eq (\ref{eq:F}) is
\beq
x_{i,k,n+1} = x_{i,k,n} + a_{n+1/2}^2\dot x_{i,k,n+1/2}/F^+_n.\label{eq:fwd2}
\eeq
Equations (\ref{eq:fwd1}) and (\ref{eq:fwd2}) are equivalent to eq. (\ref{eq:16}) at $S_{i,k,n}=0$. 

The difference from a conventional leapfrog integration is the boundary conditions, which here are the present positions and a condition on the initial velocities, as follows. 

\bigskip
A.1.4. {\it Initial conditions}

The representation of the mass distribution in the early universe by galaxy-size particles certainly is crude, but a reasonably useful approximation that motivates the following consideration. 
 
In linear perturbation theory for a continuous pressureless fluid the unwanted decaying mode has peculiar velocity that is decreasing as $v=a\dot x\propto 1/a(t)$. Eq. (\ref{eq:5}) formally eliminates this decaying mode. In the wanted growing mode the coordinate position of a particle is changing with time as 
\beq
x(t) - x(0)\propto t^{2/3}\propto a(t).
\label{eq:6}
\eeq

This implies that in the wanted growing mode the left hand side of eq. (\ref{eq:3}) is constant, meaning  $a^2dx_{i,k}/dt \simeq \hbox{ constant}\times t$, where $t$ is the time measured from $a=0$. This motivates approximating equation~(\ref{eq:3}) at the time $t_{3/2}$ intermediate between the first two time steps in the leapfrog, $a_1$ and $a_2$, as 
\beq
a_{3/2}^2{dx_{i,k,3/2}\over dt} \simeq a_{3/2}^2\dot a_{3/2}{x_{i,k,2} - x_{i,k,1}\over a_2 - a_1}= {t_{3/2}\over a_1}\left[ \sum_{j} Gm_j{(x_{j,k,1}-x_{i,k,1})\over |x_i-x_j|^3} +{1\over 2}\Omega H_o^2x_{i,k,1}\right],
\label{eq:7}
\eeq
Recall that the half time step earlier than $a_1$ is at $a_{1/2}=0=t_{1/2}=0$, where $a^2\dot x$ is supposed to vanish. Equation (\ref{eq:7}) agrees with equations (\ref{eq:bdyconds}) and (\ref{eq:16}) at $S_{i,k,1}=0$. 

It will be noted that $a_1$ may be much larger than $a_2-a_1$, meaning the numerical solution commences at modest redshift with small time steps. But $t_{3/2}$ is still the time from $a_{1/2}=0$ to the time midway between $a_1$ and $a_2$. 

The prescription in eq. (\ref{eq:7}) allows large peculiar velocities at high redshift. This is not inconsistent with eq. (\ref{eq:5}); it corresponds to a large primeval departure from homogeneity. It does mean that one must select solutions that are judged to have realistic initial peculiar velocities (at $a_{3/2}$). 

To summarize, the usual initial conditions --- position and velocity --- in a leapfrog integration of the equation of motion forward in time are replaced by the present position, $x_{i,k,n_x+1}$, and the relation in equation~(\ref{eq:7}) between the two earliest positions, $x_{i,k,1}$ and $x_{i,k,2}$, at times $t_1$ and $t_2$. If equation~(\ref{eq:6}) is a good approximation this is equivalent to specifying the initial velocity, for then eq. (\ref{eq:7}) determines $\dot x_{3/2}$. That is, the boundary conditions for a solution $S_{i,k,n}=0$ are the present position and the time-variation of the initial velocity. 

\bigskip
A.2. {\bf Method of solution}

A.2.1. {\it Single orbit adjustment}

Since we're adjusting only the orbit of particle $i$ drop  the label $i$ and write the first derivative of $S$ as
\beq
{\partial S\over\partial x_{k,n}} = S_{k,n} = -F^+_n(x_{k,n+1}-x_{k,n}) 
+  F^-_n(x_{k,n}-x_{k,n-1}) 
+ {dt_n\over a_n}\left(\sum_{j\not= i}g_{k,n;j} +
{\Omega H_o^2\over 2} x_{k,n}\right),  \label{eq:22}
\eeq
and write the equation to be solved as 
\beq
S_{k,n} + \sum_{k',n'} S_{k,n,k',n'}\delta x_{k',n'} = 0, \qquad 
{\partial^2 S\over\partial x_{k,n}\partial x_{k',n'}} = S_{k,n;k',n'} = S_{k',n';k,n}.
\label{eq:26}
\eeq
The nonzero second derivatives are
\beq
S_{k,n;k,n+1} = - F^+_n, 
\eeq
\beq
S_{k,n,k',n} =
(F^+_n + F^-_n)\delta_{k,k'} + {dt_n\over a_n}\left(\sum_{j\not= i}{\cal G}_{k,n;jk'} + {\Omega H_o^2\over 2}\delta_{k,k'}\right). 
\eeq
Since
\beq
S_{k,n;k',n'} = 0 \hbox{ unless } n'=n \hbox{ or else } n' = n\pm 1 \hbox{ and } k' = k,\label{eq:27}
\eeq
eq. (\ref{eq:26}) is
\beq
S_{k,n} + S_{k,n;k, n+1}\delta x_{k,n+1}+ 
 \sum_{k'}S_{k,n;k',n}\delta x_{k',n}
+ S_{k,n;k,n-1}\delta x_{k,n-1}= 0.  \label{eq:28}
\eeq
Set $n\rightarrow n - 1$ in this equation and rearrange it to 
\beq
\delta x_{k,n} = - {S_{k,n-1}+ \sum_{k'} S_{k,n-1;k',n-1}\delta x_{k',n-1} 
+ S_{k,n-1;k,n-2}\delta x_{k,n-2}
\over S_{k,n-1;k,n} }. \label{eq:29}
\eeq
This gives $\delta x_{k,n}$ in terms of $\delta x_{k,n-1}$ and $\delta x_{k,n-2}$. On iterating we get the form
\beq
\delta x_{k,n} = A_{k,n} + \sum_{k\pp} B_{k,n;k\pp}\delta x_{k\pp,1}.\label{eq:30}
\eeq
At $n=1$ this is just
\beq
 A_{k,1} = 0,\qquad B_{k,1;k\pp} = \delta_{k,k\pp}.
\eeq
Here and in Equation (25) the Kronecker delta $\delta$ is to be distinguished from the position shifts $\delta x$ introduced in Equation (7).
At the second time step from the start, $n=2$, eq. (\ref{eq:29}) is
\beq
\delta x_{k,2} = - \bigg[S_{k,1}+ \sum_{k'} S_{k,1;k',1}\delta x_{k',1} \bigg]/
 S_{k,1;k,2} , \label{eq:32}
\eeq
because there is no $\delta x_{k,0}$. Comparing this with eq. (\ref{eq:30}) we see that
\beq
A_{k,2} = - S_{k,1}/S_{k,1;k,2},
\qquad B_{k,2;k'} = - S_{k,1;k',1}/S_{k,1;k,2}.
\label{eq:33}
\eeq
At $n\geq 3$ the result of substituting the form (\ref{eq:30}) into eq. (\ref{eq:29})  is
\beqa
\delta x_{k,n} = &-& \bigg[S_{k,n-1} + \sum_{k'} S_{k,n-1;k',n-1}(A_{k',n-1} 
+ \sum_{k\pp} B_{k',n-1;k\pp}\delta x_{k\pp,1}) \nonumber\\
&+& S_{k,n-1;k,n-2}(A_{k,n-2} + \sum_{k\pp} B_{k,n-2;k\pp}\delta x_{k\pp,1})\bigg]
/ S_{k,n-1;k,n}. \label{eq:34}
\eeqa
So at $3\leq n\leq n_x$
\beqa
&&A_{k,n} = - {S_{k,n-1} + \sum_{k'} S_{k,n-1;k',n-1}A_{k',n-1}
+ S_{k,n-1;k,n-2}A_{k,n-2}\over S_{k,n-1;k,n}},\nonumber \\
&& B_{k,n;k\pp} = - {\sum_{k'} S_{k,n-1;k',n-1}B_{k',n-1;k\pp}
+ S_{k,n-1;k,n-2}B_{k,n-2;k\pp}\over S_{k,n-1;k,n}}. \label{eq:35}
\eeqa
Equation (A34), with Equations (A30) and (A32), gives the $A_{k,n}$ and $B_{k,n;k\pp}$ for all time steps in terms of the input derivatives of the action.  We use these to find the position shifts $\delta x_{k,n}$ in Equation (A29).  All that remains before this can be done is to compute the $\delta x_{k,1}$, which we find by setting $n=n_x$ in equation (A27) and recalling that $\delta x_{k,n_x+1}= 0$:
\beqa
0 &=& S_{k,n_x}  +  \sum_{k'}S_{k,n_x;k',n_x}\delta x_{k',n_x}
+ S_{k,n_x;k,n_x-1}\delta x_{k,n_x-1} \nonumber\\
&=&  S_{k,n_x}  +  \sum_{k'}S_{k,n_x;k',n_x}
\bigg[ A_{k',n_x} + \sum_{k\pp} B_{k',n_x;k\pp}\delta x_{k\pp,1}\bigg] \\
&& \qquad\ + S_{k,n_x;k,n_x-1}\bigg[
A_{k,n_x-1} + \sum_{k\pp} B_{k,n_x-1;k\pp}\delta x_{k\pp,1}\bigg]. \nonumber
\eeqa
So write this as
\beqa
&&0 = T_k + \sum_{k'}T_{k,k\pp}\delta x_{k\pp,1},\nonumber\\ 
&& T_k = S_{k,n_x}  +  \sum_{k'}S_{k,n_x;k',n_x}A_{k',n_x} 
+ S_{k,n_x;k,n_x-1}A_{k,n_x-1}, \label{eq:37}\\
&& T_{k,k\pp} = \sum_{k'}S_{k,n_x;k',n_x}B_{k',n_x;k\pp}
+ S_{k,n_x;k,n_x-1} B_{k,n_x-1;k\pp},\nonumber 
\eeqa
solve this $3\times 3$ set of equations for the $\delta x_{k,1}$, and then get the rest of the $\delta x_{k,n}$ from eq. (\ref{eq:30}).

A.3. {\bf Redshift boundary condition}

\smallskip

Solutions to the equations of motion in the method outlined in Appendix A.2 above satisfy the constraint that the distances at the present epoch are fixed.  However, since redshifts are known more accurately than the distances, and since we may want to explore a larger space of solutions, it may desirable to recast the problem with fixed redshifts, with the distances at the present epoch emerging as predictions.  (note that the choice of boundary condition can be made particle-by-particle.)
This can be accomplished through a partial transformation of coordinates that exchanges radial distances with radial velocities while leaving the angular position coordinates unchanged.  Details are given in Phelps (2002); the procedure is summarized below.  

The change to radial velocity coordinates is carried out in the Hamiltonian frame through a canonical transformation of the conjugate variables (positions and momenta).  The generating function of the transformation, which is added to the action outside the integral, is
\beq \label{eq:gf}
- p_{i,r,nx+1} q_{i,r,nx+1},
 \eeq
where $q_r$ and $p_r$ are the conjugate distance and momentum in the radial direction relative to the reference galaxy.  With the addition of the generating function, the problem becomes equivalent to one expressed with a new set of conjugate coordinates $Q_r$ and $P_r$, where $Q_r$ is the radial momentum and $P_r$ is the radial distance.  In these coordinates the boundary term in the variational derivative $P \delta Q$ vanishes at $z=0$ when the {\it angular} positions and the {\it radial} velocity vanish.  However, since the transformed Hamiltonian cannot be expressed analytically (the gravitational term cannot be written out in terms of $Q$), the computation must be carried out in the original coordinate system, expressed in the Lagrangian frame, where the new boundary condition must be imposed by hand to recover the correct equations of motion.  That constraint takes the form of additional terms in the action, as follows.

A.3.1 {\it Modification to the action}

The modified action is:
\beqa
S &=& {1 \over 2} H_0 x_{i,r,nx+1}^2 - cz x_{i,r,nx+1} + x_{i,r,nx+1} \vec{v}_{mw,nx+1} \cdot \hat{r}_{i,nx+1} \nonumber \\
&+&\sum_{i,k,n=1,n_x}{1\over 2}{(x_{i,k,n+1}- x_{i,k,n})^2\over (a_{n+1}-a_n)}
\dot a_{n+1/2}a_{n+1/2}^2 \nonumber\\
&+&\sum_{i,j,n=1,n_x}{t_{n+1/2}-t_{n-1/2}\over a_n}
\left[\sum_{j<i}{Gm_j\over  |x_{i,n}-x_{j,n}|} + {1\over 4}
\sum_i \Omega H_o^2 x_{i,n}^2\right].
\eeqa

The gradient at the final time step is:
\beqa
S_{i,k,n} &=& H_0 x_{i,r,nx+1} - cz + \vec{v}_{mw,nx+1} \cdot \hat{r}_{i,nx+1} 
\nonumber \\
& & -F^+_n(x_{i,k,n+1}-x_{i,k,n}) +  F^-_n(x_{i,k,n}-x_{i,k,n-1})
+ {dt_n\over a_n}\left[g_{i,k,n}  + 
{1\over 2}\Omega H_o^2 x_{i,k,n}\right].
\eeqa

The modified second derivatives of the action at the final two timesteps are:
\beq
S_{i,k,nx;i,k,nx+1} = - {\hat{x}_{i,k,nx+1} \over dt_{nx+1}} 
\eeq
Note that a new term in the action $S_{k,n_x+1;k,n_x+1}$ is also present, but it is not used in the computation that follows.

Since the boundary condition is now velocity-limited, $\delta x_{k,n_x+1}\not= 0$, and we must add a new term in eq. (\ref{eq:gf}) arising from the final two timesteps:
\beqa
0 &=& S_{k,n_x}  +  \sum_{k'}S_{k,n_x;k',n_x}\delta x_{k',n_x} 
+ S_{k,n_x;k,n_x-1}\delta x_{k,n_x-1}    \nonumber  \\
&& \qquad +  S_{k,n_x;k,n_x+1}\delta x_{k,n_x+1}  \nonumber \\
&=&  S_{k,n_x} +  \sum_{k'}S_{k,n_x;k',n_x}
\bigg[ A_{k',n_x} + \sum_{k\pp} B_{k',n_x;k\pp}\delta x_{k\pp,1}\bigg]  \nonumber \\
&& \qquad\ + S_{k,n_x;k,n_x-1}\bigg[
A_{k,n_x-1} + \sum_{k\pp} B_{k,n_x-1;k\pp}\delta x_{k\pp,1}\bigg]   \label{eq:sk} \\
&& \qquad\  +  S_{k,n_x;k,n_x+1}
\bigg[ A_{k,n_x+1} + \sum_{k\pp} B_{k,n_x+1;k\pp}\delta x_{k\pp,1}\bigg].  \nonumber
\eeqa
This is now written as
\beqa
&&0 = T_k + \sum_{k'}T_{k,k\pp}\delta x_{k\pp,1}, \nonumber \\ 
&& T_k = S_{k,n_x}  +  \sum_{k'}S_{k,n_x;k',n_x}A_{k',n_x} 
+ S_{k,n_x;k,n_x-1}A_{k,n_x-1},  \nonumber \\ 
&& \qquad\ + S_{k,n_x;k,n_x+1}A_{k,n_x+1},  \label{eq:tk} \\
&& T_{k,k\pp} = \sum_{k'}S_{k,n_x;k',n_x}B_{k',n_x;k\pp}
+ S_{k,n_x;k,n_x-1} B_{k,n_x-1;k\pp}, \nonumber \\
&& \qquad\ + S_{k,n_x;k,n_x+1}B_{k,n_x+1;k\pp}.  \nonumber 
\eeqa

As before, this $3\times 3$ set of equations is solved for the $\delta x_{k,1}$, and we get the rest of the $\delta x_{k,n}$ from eq. (\ref{eq:30}).  However, we will find below that the boundary conditions will enable us to replace the $\delta x_{nx+1}$ above with functions of $\delta x_{nx}$.

A.3.2 {\it Modification to the position shifts}

The change of boundary conditions implies a relationship between the position shifts $\delta x_{k,nx+1}$ and $\delta x_{k,nx}$ that will further modify the new terms above.  The relationship is simplest to see in the one-dimensional case (setting $y = z = 0$), where
\beqa
cz &=&  v_{nx+1} + H_0 x_{nx+1} - v_{mw,nx+1}. \nonumber \\ 
 &=&  {x_{nx+1} - a_{nx} x_{nx} \over dt_{nx+1/2}} + H_0 (x_{nx+1} - x_{mw,nx+1}) - v_{mw,nx+1}, \label{eq:cz0}
\eeqa
where we continue to drop the particle subscript unless referring to the reference galaxy, $mw$, since we are working on one particle at a time while holding the rest of the catalog fixed.   

After the variation,
\beqa
cz' &=&  {x_{nx+1} + \delta x_{nx+1} - a_{nx} x_{nx} - a_{nx} \delta x_{nx} \over  dt_{nx+1/2}} \nonumber \\ 
 & & + H_0 (x_{nx+1} + \delta x_{nx+1} - x_{mw,nx+1}) - v_{mw,nx+1}.
\eeqa
Recall that the orbit of the reference galaxy, and thus its contribution to the redshift, is held constant during the variation.

Setting $cz' = cz$ gives
\beqa
\delta x_{nx+1}  &=& {a_{nx} \delta x_{nx} \over 1 + H_0 dt_{nx+1/2}}.
\eeqa

In three dimensions the angular coordinates of the particle must also remain unchanged after the variation, and so now there are three equations constraining the coordinate shifts at the final two timesteps:
\beq
cz' = cz ,\qquad   \theta' = \theta,\qquad \phi' = \phi,
\eeq
where $\theta$ and $\phi$ are defined relative to the reference galaxy at $z=0$:
\beqa
\tan \theta_{nx+1}  &=& {x_{2,nx+1} - x_{mw,2,nx+1} \over x_{1,nx+1} - x_{mw,1,nx+1}} \\
\tan \phi_{nx+1} &=& {x_{3,nx+1} - x_{mw,3,nx+1} \over \sqrt{(x_{1,nx+1} - x_{mw,1,nx+1})^2 + (x_{2,nx+1} - x_{mw,2,nx+1})^2 }}.
\eeqa 

The $\theta$ and $\phi$ constraints impose a relationship between the three $\delta x_{k,nx+1}$.  The $\theta$ constraint gives:
\beqa
\theta'_{nx+1} &=& \theta_{nx+1} \nonumber \\ 
{x'_2 \over x'_1} &=& {x_2 \over x_1} \nonumber \\
{x_{2,nx+1} - x_{mw,2,nx+1} + \delta x_{2,nx+1} \over x_{1,nx+1} - x_{mw,1,nx+1} + \delta x_{1,nx+1}} &=& {x_{2,nx+1} - x_{mw,2,nx+1} \over x_{1,nx+1} - x_{mw,1,nx+1}} \nonumber \\
 \delta x_{2,nx+1}  &=& {x_{2,nx+1} \over x_{1,nx+1}} \delta x_{1,nx+1}.
\eeqa
The combined $\theta$ and $\phi$ constraints further give:
\beqa
\phi'_{nx+1} &=& \phi_{nx+1}  \nonumber \\
{x_1'^2 + x_2'^2 \over x_3'^2} &=& {x_1^2 + x_2^2 \over x_3^2} \nonumber \\
\delta x_{3,nx+1}  &=& {x_{3,nx+1} \over x_{1,nx+1}} \delta x_{1,nx+1}.
\eeqa
As for the third constraint, the redshift in three dimensions is:
\beqa \label{eq:cz}
cz &=& (\vec{v}_{nx+1} - \vec{v}_{mw,nx+1}) \cdot \hat{r}_{nx+1}  + H_0 r_{nx+1}  \nonumber \\
  &=& \sum_k \bigg( {x_{k,nx+1} - a_{nx} x_{k,nx} \over dt_{nx+1/2}} - \vec{v}_{mw,nx+1} \bigg) \cdot \hat{x}_{k,nx+1}  \nonumber \\
 & & + H_0 \sqrt{\sum_k (x_{k,nx+1} - x_{mw,k,nx+1})^2},
\eeqa
where
\beq
\hat{x}_{k,nx+1} = {x_{k,nx+1} \over r_{nx+1}} = {x_{k,nx+1} \over \sqrt{\sum_{k'} (x_{k',nx+1} -x_{mw,k',nx+1} )^2}}.
\eeq
After the position shifts,
\beqa
cz' &=& \sum_k \bigg( {x_{k,nx+1} + \delta x_{k,nx+1} - a_{nx} x_{k,nx} - a_{nx} \delta_{k,nx} \over dt_{nx+1/2}} - \vec{v}_{mw,k,nx+1} \bigg) \cdot \hat{x}'_{k,nx+1}  \nonumber \\
 & &+ H_0 \sqrt{\sum_k (x_{k,nx+1} - x_{mw,k,nx+1} + \delta x_{k,nx+1}) ^2}
\eeqa
where 
\beqa
\hat{x}'_{k,nx+1} &=& {x_{k,nx+1} + \delta x_{k,nx+1} \over \sqrt{\sum_{k'} (x_{k',nx+1} - x_{mw,k',nx+1} + \delta x_{k',nx+1})^2}}.
\eeqa

Setting $cz' = cz$ and using the approximation $\delta x_k / x_k \ll 1$, we find, after some algebra,
\beq \label{eq:dx}
\delta x_{k,nx+1} = {x_{k,nx+1} \over \alpha} \sum_{k'} x_{k',nx+1} a_{nx} \delta x_{k',nx},
\eeq
where
\beqa
\alpha &=& \sum_{k\pp} x_{k\pp,nx+1} \bigg(x_{k\pp,nx+1} (2 -\beta) - a_{nx} x_{k\pp,nx} - v_{mw,k\pp,nx+1} dt_{nx+1/2}\bigg) \\
\beta &=& 1 - H_0 dt_{nx+1/2} - \sum_{k'''} {x_{k''',nx+1} a_{nx} x_{k''',nx} + x_{k''',nx+1} v_{mw,''',nx+1} dt_{nx+1/2} \over r_{nx+1}^2}.
\eeqa
It can be shown that this reduces to the one-dimensional case when we set two of the three position coordinates to zero.

With the above, the position shifts $\delta x_{k,nx+1}$ can be replaced in eqs. (\ref{eq:sk}-\ref{eq:tk}) by their equivalent expressions in terms of $\delta x_{k,nx}$:
\beqa
0 &=&  S_{k,n_x} +  \sum_{k'}S_{k,n_x;k',n_x}
\bigg[ A_{k',n_x} + \sum_{k\pp} B_{k',n_x;k\pp}\delta x_{k\pp,1}\bigg]  \nonumber \\ 
&& \qquad\ + S_{k,n_x;k,n_x-1}\bigg[
A_{k,n_x-1} + \sum_{k\pp} B_{k,n_x-1;k\pp}\delta x_{k\pp,1}\bigg] \\ 
&& \qquad\  +  S_{k,n_x;k,n_x+1}
x_{k,nx+1} \alpha^{-1} \sum_{k'} x_{k',nx+1} \bigg[ A_{k',n_x} + \sum_{k\pp} B_{k',n_x;k\pp}\delta x_{k\pp,1}\bigg].  \nonumber
\eeqa
\beqa
&&0 = T_k + \sum_{k'}T_{k,k\pp}\delta x_{k\pp,1}, \nonumber \\ 
&& T_k = S_{k,n_x}  +  \sum_{k'}S_{k,n_x;k',n_x}A_{k',n_x} 
+ S_{k,n_x;k,n_x-1}A_{k,n_x-1}, \nonumber \\ 
&& \qquad\ + S_{k,n_x;k,n_x+1} x_{k,nx+1} \alpha^{-1} \sum_{k'} x_{k',nx+1} A_{k',n_x} \\ 
&& T_{k,k\pp} = \sum_{k'}S_{k,n_x;k',n_x}B_{k',n_x;k\pp}
+ S_{k,n_x;k,n_x-1} B_{k,n_x-1;k\pp}, \nonumber \\
&& \qquad\ + S_{k,n_x;k,n_x+1} x_{k,nx+1} \alpha^{-1} \sum_{k'} x_{k',nx+1} B_{k',n_x;k\pp} \nonumber
\eeqa

Note that the actual value of the redshift has been nowhere specified and must be arranged by hand, as by choosing the initial trial orbits to have the input redshifts.  However, the approximation $\delta x_k / x_k \ll 1$ used to arrive at eq. (\ref{eq:dx}) can fail, particularly in the first few iterations of the action, and so in practice the redshifts may drift away from their input values as the stationary point in the action is reached.  This drift is corrected by periodically adjusting the particle positions at timestep $nx+1$ according to eq. (\ref{eq:cz}).

\end{document}